\documentstyle[12pt,epsfig]{article}
\newcommand{\be}{\begin{equation}}
\newcommand{\ee}{\end{equation}}
\newcommand{\bear}{\begin{eqnarray}}
\newcommand{\eear}{\end{eqnarray}}
\newcommand{\ba}{\begin{array}}
\newcommand{\ea}{\end{array}}

\def\aa{\alpha}

\newskip\humongous \humongous=0pt plus 1000pt minus 1000pt

\newif\ifdtup

\def\oldreffmt#1{\rlap{[#1]} \hbox to 2\parindent{}}

\def\figfmt#1{\rlap{Figure {#1}} \hbox to 1in{}}

\def\slash#1{#1\!\!\!/\!\,\,}
\def\beq{\begin{equation}}
\def\eeq{\end{equation}}
\def\bea{\begin{eqnarray}}
\def\eea{\end{eqnarray}}
\def\half{\frac{1}{2}}

\def\bq{\begin{quote}}
\def\eq{\end{quote}}

\def\half{\frac{1}{2}}     
\def \lta {\mathrel{\vcenter
     {\hbox{$<$}\nointerlineskip\hbox{$\sim$}}}}
\def \gta {\mathrel{\vcenter
     {\hbox{$>$}\nointerlineskip\hbox{$\sim$}}}} 

\relax

\newdimen\tdim
\tdim=\unitlength
\def\bar{\overline}

\begin{document}

\pagestyle{empty}
\begin{titlepage}
\def\thepage {}    

\title{  \bf  
Gauge Invariant Effective Lagrangian \\[2mm]
for Kaluza-Klein Modes \\[10mm] } 
\author{
\bf  Christopher T. Hill$^1$ \\[2mm]
\bf Stefan Pokorski$^{1,2}$ \\[2mm]
\bf Jing Wang$^1$ \\ [2mm]
 {\small {\it $^1$Fermi National Accelerator Laboratory}}\\
{\small {\it P.O. Box 500, Batavia, Illinois 60510, USA}}
\thanks{e-mail: hill@fnal.gov, jingw@fnal.gov,
Stefan.Pokorski@fuw.edu.pl}\\
{\small {\it $^2$ Institute for Theoretical Physics}}\\
{\small{\it Hoza 69, 00-681, Warsaw, Poland}} 
}

\date{\today}

\maketitle

\vspace*{-14.0cm}
\noindent

\begin{flushright}
FERMILAB-Pub-01/043-T \\ [1mm]
October, 2000
\end{flushright}

\vspace*{14.1cm}
\baselineskip=18pt

\begin{abstract}

  {\normalsize
We construct a manifestly gauge invariant Lagrangian 
in $3+1$ dimensions for $N$ Kaluza-Klein modes
of an $SU(m)$ gauge theory in the bulk. For example,
if the bulk is $4+1$, the effective theory is
$\Pi_{i=1}^{N+1} SU(m)_i$ with $N$ chiral $(\bar{m},m)$
fields connecting the groups sequentially. 
This can be viewed
as a Wilson action
for a transverse lattice in $x^5$, and is shown 
explicitly to match the continuum $4+1$ compactified Lagrangian 
truncated in momentum space.
Scale dependence of the gauge couplings is described by the 
standard renormalization group technique with threshold
matching, leading to effective power law running. We also discuss
the unitarity constraints, and chiral fermions.  
} 
\end{abstract}

\vfill
\end{titlepage}

\baselineskip=18pt
\renewcommand{\arraystretch}{1.5}
\pagestyle{plain}
\setcounter{page}{1}


\section{Introduction}

It is widely believed
that the main low energy signature of extra
dimensions 
is the appearance of the tower of Kaluza-Klein (KK) modes
\cite{largeextradimensions}. For example,
if QCD lived in the bulk, experimentalists would see
massive spin-1 degenerate color octet vector bosons (colorons)
appearing at large mass scales corresponding to (inverse)
compactification scales.
As these new massive KK particles begin to emerge in 
accelerator
experiments, 
we might ask how would we describe
them in an effective four-dimensional renormalizeable
Lagrangian that is an extension of QCD,
without an {\em a priori} knowledge of the existence of extra dimensions? 
The main goal of the present paper is to give a manifestly
gauge invariant effective Lagrangian description of KK modes
in $3+1$ dimensions.

It is important to realize
at the outset that 
there is an implicit dynamical assumption underlying a theory
with extra-dimensions and KK modes.  This is the assumption
that there is a meaningful separation of scales between
the compactification scale, $M_c \sim 1/R$ and the
``string'' or  ``fundamental scale''
$M_s$ at which the extra-dimensional theory breaks down 
as a perturbative local field theory.  
To have $N>>1$ KK modes in a $4+1$
theory we require $M_s/M_c \sim N >> 1$.  
It is not obvious how such a
separation of scales occurs in the theory
(It involves soft mass scales in the radion potential
that somehow remain isolated from $M_s$).
Can it occur naturally or does it
require fine-tuning?   Such a
hierarchy requires strong
coupling at the high energy scale
$M_s$.   We will assume, as
do all extra-dimensional models, that we
have such a hierarchy, and return to this
issue in Section 6.

Having engineered a hierarchy with
$N>>1$ KK modes, by analogy with critical behavior 
in a second order phase transition in condensed matter physics,
there should exist a wide
range, or universality class, 
of theories that have identical behavior in the infra-red, but
are radically different in detail at the scale $M_s$.  
In the present paper we exploit universality.  
We treat the physics at $M_s$ not as
a ``string theory,'' but rather as a ``transverse
lattice gauge theory'' \cite{transverse}.
For us, the normal $3+1$ dimensions of space-time are
continuous, but the extra dimensions are latticized
(nothing prevents us from adopting a full lattice theory,
but it is convenient for our presnt purposes to use the transverse
lattice).
This theory will have a well-defined finite short-distance
behavior for arbitrarily large
coupling and will be manifestly gauge invariant, reflecting
the full gauge invariance of the higher dimensional theory.
It will have the same infra-red behavior as the usual
KK-mode description, but will illuminate how the gauge
invariance is maintained.

As a result, we understand something implicitly puzzling
about KK modes.  Longitudinal KK mode scattering is
essentially the scattering of Nambu-Goldstone bosons 
in a nonlinear chiral Lagrangian. As such it 
violates perturbative unitarity, i.e., there is
a Lee-Quigg-Thacker bound on the applicability
of the theory \cite{Lee}.  We will see that this
happens at, none other than,
the scale $M_s$ in our effective Lagrangian.
This is not surprising, since the parent $D=5$ theory
has a dimensional coupling constant $g_0$ 
(with dimension $M^{-1/2}$) and is
expected to violate perturbative unitarity
when $s \gta M_s/\alpha_0$. 
This indeed translates into the unitarity 
bound $s \lta 4\pi v^2$ for longitudinal gauge boson scattering in our
effective $3+1$ theory.

The main reason for desiring an approach such as this is that
it is difficult to treat nonabelian gauge theories 
in loop expansions with momentum space cut-offs. Normally,
the momentum space cut-off is not compatible with gauge invariance,
and this causes the loop expansion to become non-gauge invariant.
However, the usual treatment of extra-dimensional gauge theories
involves a truncation on KK modes, which is a de facto
momentum space cut-off. 
With gauge fields in the bulk, a $d+1$ theory with $d>3$
has
infinitely more gauge invariance than the $3+1$
theory since there is more space in which
to perform local gauge transformations. 
Clearly the gauge invariance of $3+1$ QCD must be
maintained, but how does the expanding local gauge invariance
of the theory manifest itself as the
extra dimension begins to open up with the 
emergence of KK modes?  How does the power-law
running of the coupling constant emerge and what is
the correct renormalization group for such
a description?

\section{Manifestly Gauge Invariant Effective Lagrangian} 
 
The KK modes of the vector bosons of QCD,
i.e., the colorons, are heavy matter fields and must
transform linearly under the adjoint representation of $SU(3)$
(in contrast to the zero-mode gluon which transforms nonlinearly
by the Yang-Mills gauge transformation).
References \cite{hidden} have argued 
that vector fields in linear adjoint representations 
of a local gauge group $SU(m)$ will
always contain a ``hidden'' local symmetry, which is a copy of
$SU(m)$.
The gluon plus one massive octet vector multiplet 
corresponds to the local symmetry 
$SU(m)\times SU(m)$, each factor having
the same coupling constant (our present discussion is
classical; we'll worry about running couplings below). 
This is broken diagonally by an effective Higgs  field, $\Phi$,
which transforms as a $(\bar{m},m)$, to a local $SU(m)$
and an $SU(m)$ global symmetry. 
Only the chiral
components of $\Phi$ are relevant here
so we can replace $\Phi \rightarrow v \exp(\phi^a\lambda^a/2v)$
(see footnote [1]).  
The $\phi^a$ are eaten to give the coloron mass.   
Hence, in describing one massive octet 
this way it is the low energy hidden local symmetry due
to the spontaneous breaking
that reflects the 
expanded gauge invariance of the extra-dimensional
theory as the space of the extra dimension is opening up.

As experiments go to higher energies, one starts to see
more KK massive gauge bosons. It is obvious that one requires more ``hidden''
local $SU(3)$ symmetries and more Higgs fields as in the previous case 
to construct an effective Lagrangian to
describe these massive gauge bosons. Hence, we propose that the effective
Lagrangian for the first $n$ KK modes would contain $N+1$ 
($N \gg n$) $SU(3)$'s with $N$
$\Phi$'s.  The interconnections between the gauge symmetries and the Higgs
could become completely arbitrary, and resolve into different hydrocarbon-like
chain molecules.  

%

We might guess that the simplest linear interconnection for $N$ modes
having $\Phi_i\subset (\overline{3}_i,3_{i+1})$ is somehow
relevant. We'll follow the organic chemistry nomenclature and call
this an ``aliphatic'' $(SU(3)^{N+1},\Phi^{N})$ 
model. The Lagrangian for this scheme
is:\footnote{\noindent
A renormalizable potential can be constructed for the Higgs fields, 
\be
V(\Phi_j) = \sum_{j=1}^{N} \left[ -M^2 {\rm Tr}(\Phi_j^2) + \lambda_1 {\rm
Tr}(\Phi_j^4) + 
\lambda_2 {\rm Tr}(\Phi_j^2)^2 + M^{'} det(\Phi_j) \right],
\label{V} 
\ee
We can always  
arrange the parameters in the potential such that the diagonal
components of each $\Phi_j$ develop a vacuum 
expectation value $v$, and the Higgs and $U(1)$ PNGB are heavy.
}
\be
{\cal{L}}= -\frac{1}{4}\sum_{i=0}^N F_{i\mu\nu}^a F^{i\mu\nu a} +
\sum_{i=1}^{N} D_\mu\Phi_i^\dagger D^\mu\Phi_i
\ee
in which the covariant derivative is defined as $D_{\mu}= \partial_{\mu} + i
g_L A_{\mu}^{a}T^{a}$, $g_L$ is the dimensionless gauge coupling constant 
that is equal for all of
the $SU(3)$ symmetries and $T^{a}$ are the generators of
the gauge symmetry where $a$ is the color index. 
Note that the fact that $g_L$ is common
for all the gauge groups is a key constraint and 
would be to the experimentalist in $3+1$ evidence
of the extra-dimensions.  
Upon substituting,
\be 
\Phi_i \rightarrow v\exp(i\phi^a_i\lambda^a/2v)
\ee
the $\Phi$ kinetic terms lead to a mass matrix for the gauge
fields:
\be
\sum_{i=1}^{N} \half g_L^2v^2(A^a_{(i-1)\mu} -A^a_{i\mu})^2
\ee
This mass matrix has the structure of a nearest neighbor 
coupled oscillator
Hamiltonian. We can diagonalize the mass
matrix to find the eigenvalues (which corresponds to the
dispersion relation for the coupled oscillator-system):
\be
\label{Mn}
M_n = \sqrt{2}g_Lv \sin \left[ \frac{\gamma_n}{2} \right]
\qquad \gamma_n=
\frac{n\pi}{N+1}\ , \qquad n=0,1,\dots, N. 
\ee
Thus we see that for small $n$ this system has a KK tower
of masses given by:
\be
M_n \approx   \frac{g_Lv\pi n}{\sqrt{2}(N+1)}\qquad \qquad n << N
\ee
and $n=0$ corresponds to the zero-mode gluon. 

To match on to the spectrum of the KK modes, we require 
\be
\frac{g_Lv}{\sqrt{2}(N+1)} = \frac{1}{R}. 
\ee
Hence, the aliphatic system with $SU(3)^{N+1}$ and $N$ $\Phi_i$ provides 
a gauge invariant description of the first $n$ KK modes by generating the same 
mass spectrum. It is thus crucial to examine the interactions from the
aliphatic model.

In a geometric picture, the aliphatic model corresponds to a
``transverse lattice'' description of a full $4+1$
gauge theory \cite{transverse}.
We construct a transverse lattice in the $x^5$ dimension
where the lattice size is $R$ and short-distance lattice cut-off
is $a$, so $N =R/a$. This
is a foliation of $N+1$ parallel branes, each spaced
by a  lattice cut-off $a$ 
(Fig.(1)).
\begin{figure}
\vskip -1in
\centerline{
\hbox{
\epsfxsize=2.5truein
\epsfbox[70 32 545 740]{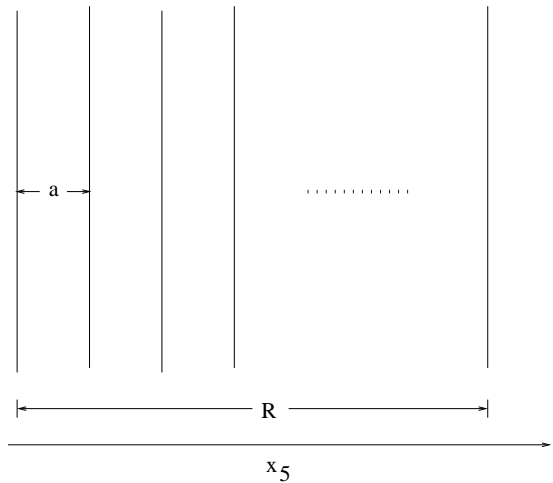}
}
}
\caption{The geometric interpretation for the aliphatic model as a
transverse lattice in the $x^5$ dimension with continuum
theory in $4+1$.  The number of
branes in the foliation is $N+1 = (R/a) + 1$. }  
\end{figure}
 On the $i$th brane
we have an $SU(m)$ gauge theory denoted by $SU(m)_i$.  The
$SU(m)_i$ automatically have a common 
coupling constant $g$.
Each brane $SU(m)_i$ theory can be viewed as predefined
in the continuum limit of a fine-grained Wilson plaquette action,
and a hypothetical $3+1$ lattice spacing $a_4$.
The lattice spacing in the $x^5$ dimension can be viewed as
relatively coarse with $a >> a_4$ \cite{transverse}.

The theory thus
has $N$ links in the $x^5$ direction that are
continuous functions of $x_\mu$. These
correspond to the continuum limit Wilson lines:
\be
\Phi_n(x^\mu) = \exp\left[ig_0\int_{na}^{(n+1)a}\; dx^5 A_5(x^\mu, x^5)\right]
\rightarrow \exp\left[ig_0a A_5(x^\mu, (n+\half)a)\right]
\ee
The $N$ $\Phi_n$ therefore transform as an $(\bar{m},m)$ representation
of $SU(m)_n\times SU(m)_{n+1}$ as in the aliphatic model
(straddling the
nearest neighbor $SU(m)_n$ and $SU(m)_{n+1}$ gauge groups).
$\Phi_n$ is a unitary matrix and may be parameterized 
as in eq.(1.3). 
The theory is a spline approximation to the configurations
in the continuum $x^5$ dimension.

\newpage
\section{Compare the Continuum Theory }

\noindent
{\em (i) Definition of the Continuum Theory}
\vskip 0.2cm

A $d+1$ ($d>3$) field theory becomes ill-defined at energy scale $M_s >>
1/R$. Presumably it matches onto a string theory at $M_s$, and
we usually refer to $M_s$ as the ``string scale.'' 
While the exact structure of the theory
on scales $\mu \sim M_s$ is unknown, its symmetries, e.g., 
local gauge invariance, must remain intact at lower
scales. A continuum $d+1$ Yang-Mills
Lagrangian gives a valid description at scales below $M_s$. 

A Wilson transverse lattice Lagrangian
is a reasonable candidate for a 
well-defined short distance definition 
of the nonperturbative higher
dimensional theory. This manifestly preserves local gauge invariance
and permits, in principle, a nonperturbative treatment. 
How, then, does the aliphatic $(SU(3)^{N+1},\Phi^{N})$ model
match in detail to the 
perturbative $4+1$ continuum theory at lower energies?

We define the continuum theory in $4+1$ and expand
in modes in the compact $x^5$.  We  truncate this theory
after $N$ terms. Now, momentum space truncations in Yang-Mills theories
are notoriously awkward at best. The expansion is usually
done in a particular gauge. Then, with truncation of the theory 
in momentum space we lose track of the full gauge invariance
of the theory.  However, we will see, remarkably, 
that this truncation 
can be matched identically onto the aliphatic theory which is manifestly 
gauge invariant. 
Since the aliphatic model is manifestly gauge invariant and 
renormalizable, various field theoretical
questions can be given precise formulation. One of them is the running of the 
coupling constants, which at one loop level qualitatively agrees 
with the results of Dienes,  
et al. \cite{dienes} and Dobrescu et al. \cite{dobrescu}. But in our formulation it can be
systematically calculated to any required degree of accuracy. 

First, we consider a simple 
well-defined compactification scheme.
We define QCD in $4+1$ dimensions
between two parallel branes.\footnote{The ordinary 
spacetime coordinates are  labeled by $x^\mu$, $\mu=0,1,2,3$, 
and  the fifth dimension by $x^5$ to avoid confusion with
$x^4=ict$; Capital
letters denote the bulk coordinates, $M,N=0,1,2,3,5$.}.
The branes are respectively located at I: 
$x^5=R_I = 0 $ and II: $x^5= R_{II} = R $, with a constant 
inter-brane separation $R$.
The covariant
derivative is defined as $D_M = \partial_M + i  g_0\hat{A}_M^{a}T^{a}$,  
with  field
strengths $ig_0\hat{F}_{MN} = [D_M, D_N]$, where the canonical mass
dimension of the vector 
potential $\hat{A}_M$ in $4+1$ dimensions is $3/2$, and the coupling
constant $g_0$ must therefore have dimension
$-1/2$. 

The five-dimensional theory is locally gauge invariant but
non-renormalizable. In addition to the compactification radius $R$, it is
defined by the fundamental short-distance cut-off scale $M_s$. It is then
natural to define a dimensionless
$g$ by $g_0\equiv 1/\sqrt{M} = g/\sqrt{M_s} $. 
The $4+1$ Lagrangian takes the form:
\be
{\cal{L}}_5 = -\frac{1}{4}Tr(\hat{F}_{MN}\hat{F}^{MN}),  \qquad 
\hat{F}_{MN}^a =
\partial_M \hat{A}_N^a - \partial_N \hat{A}_M^a + g_0 f^{abc}\hat{A}_M^b \hat{A}_N^c; 
\ee
where $a$ is the gauge index and $f^{abc}$ is the structure constant.

\noindent
{\em (ii) Momentum Space Expansion and Truncation}
\vskip 0.1cm
 
A necessary gauge-covariant  boundary condition 
is:
\be 
 F^{5N} =  F^{N5} = 0\ ,\qquad {\rm at}~~ x^5= R_{I,II}
\label{bc0}
\ee
This removes  unwanted gauge invariant vector field strengths
that transform as a $4$-vector in the $3+1$ theory.
The simplest gauge
choice realizing these boundary conditions is to impose
Neumann conditions for 
$\hat{A}_\mu$ with $\mu=0,1,2,3$, i.e.
$\partial \hat{A}_\mu/ \partial x^5 = 0$, at $x_5 =
R_{I,II}$, and
Dirichlet conditions for the $3+1$ ``scalars'' $\hat{A}_5$, i.e.
$\hat{A}_5 = 0$ at $x_5 = R_{I,II}$.    The lowest energy
physical $\hat{A}_\mu$ modes are massless, independent of $x_5$, and
form the usual $3+1$ gauge field. We can further
choose an axial gauge $\chi^A\hat{A}_A=0$ where $\chi^A$ is a
5-vector normal to the branes. This sets $\hat{A}^5 =0$.
We will adopt this
gauge choice after the momentum space expansion.

We thus can expand the 4-vector
potential $\hat{A}_{\mu}(x_{\mu}, x_{5})$ in a
Fourier cosine series, 
\be
\hat{A}_{\mu}=\frac{1}{\sqrt{R}}\left[ A_{\mu}^0 +\sqrt{2}
\sum^{+\infty}_{n=1} A_{\mu}^n(x_{\mu})\cos{(n\theta)} \right]\ , \qquad \theta = \frac{\pi x_5}{R}, 
\label{An}
\ee
where we have suppressed the gauge index $a$ and $A^0$
is the $n=1$ zero-mode. The fifth component
$\hat{A}_{5}(x_{\mu}, x_{5})$ is given by a Fourier sine series,
\be
\hat{A}_{5}=\sqrt{\frac{2}{R}} \sum^{+\infty}_{n=1} A_{5}^n(x_{\mu})\sin{(n\theta)}\ . 
\label{A5n}
\ee
and this has no zero-mode.
The coefficients of the expansions are:
\begin{eqnarray}
A_{\mu}^{0} & = & \frac{1}{2\sqrt{R}} \int_{0}^{R} dx_5 \hat{A}_{M}(x_{\mu},
x_5); \\ \nonumber 
A_{\mu}^{n} & = & \frac{1}{\sqrt{2R}} \int_{0}^{R} dx_5 \hat{A}_{M}(x_{\mu}, x_5)
\cos(n\theta)\ ; \qquad n=1, \cdots, +\infty \\ \nonumber
 A_{5}^{n} & = & \frac{1}{\sqrt{2R}} \int_{0}^{R} dx_5 \hat{A}_{5}(x_{\mu}, x_5)
\sin(n\theta)\ ; \qquad n=1, \cdots, \infty.  \\ \nonumber
\end{eqnarray}
The non-hat vector field $A_{M}^n$ has mass dimension $+1$. 

The field strengths read, 
\begin{eqnarray}
\hat{F}_{\mu \nu}(x_{\aa}, x_5) & = & \frac{1}{\sqrt{R}} \left\{
\left[ \partial_{[\mu}A_{\nu]}^0 + \sum^{+\infty}_{n=1}\cos{(n\theta)}
\partial_{[\mu}A_{\nu]}^{n} \right] \right. \\ \nonumber
 & & + \left. \frac{g}{\sqrt{M_sR}}f^{abc} \left[ A_{\mu}^0 +\sqrt{2}
\sum^{+\infty}_{n=1} A_{\mu}^n\cos{(n\theta)} \right] \left[A_{\mu}^0 +\sqrt{2}
\sum^{+\infty}_{m=1} A_{\mu}^m\cos{(m\theta)} \right] \right\}, 
\label{Fmun}
\end{eqnarray}
the color indices on the vector fields are supressed in this equation as well
as in the following equations.  
Integrating over $x^5$ we obtain the effective $3+1$ theory. 

If we now impose the axial gauge $A_5(x_{\mu}, x_5) \equiv 0$, the effective
Lagrangian after integrating over $x_5$ and truncating at the $N$th KK mode
takes the form: 
\begin{eqnarray}
{\cal{L}}_{4} & = & (\partial_{\mu}A_{\nu}^0-\partial_{\nu}A_{\mu}^0 + 
\frac{g}{\sqrt{M_sR}}f^{abc} A_{\mu}^0A_{\nu}^0)^2
+ \sum_{n=1}^{N}(\partial_{\mu}A_{\nu}^n-\partial_{\nu}A_{\mu}^n)^2 
\\ \nonumber
& & +\frac{2g}{\sqrt{M_s R}} f^{abc} \sum_{n=1}^{N} \left[
\partial_{[\mu}A_{\nu]}^0 A^{n~\mu}A^{n~\nu} + \partial_{[\mu}A_{\nu]}^n(
A^{0~\mu}A^{n~\nu} + A^{n~\mu}A^{0~\nu}) \right] 
\\ \nonumber
& & + \frac{g}{\sqrt{2M_s R}}f^{abc}\sum_{n,m,l=1}^{N}
\partial_{[\mu}A_{\nu]}^n A^{m~\mu} A^{l~\nu} \Delta_1(n,m,l)
\\ \nonumber
& & + \frac{g^2}{M_s R}
f^{abc}f^{ade}\sum_{n=1}^{N}\left(A_{\mu}^0A_{\nu}^0A^{n\mu}A^{n\nu} +
\makebox{all
permutations} \right)
\\ \nonumber 
& & +
\frac{g^2}{2M_s R}f^{abc}f^{ade}\sum_{n,m,l,k=1}^{N}A_{\mu}^nA_{\nu}^mA^{l~\mu}A^{k~\nu}
\Delta_2(n,m,l,k) 
\\ \nonumber
& & + \sum_{n=1}^{N}(\frac{n\pi}{R})^2A_{\mu}^{n}A^{n\mu}  , 
\label{L4}
\end{eqnarray}
where the $\Delta_i$ are defined as:
\begin{eqnarray}
& & \Delta_1 =
\delta(n+m-l)+\delta(n-m+l)+\delta(n-m-l)
\\ \nonumber
& & \Delta_2 = \delta(n+m-l-k)+\delta(n+m+l-k)+\delta(n+m-l+k)
\\ \nonumber
& & \qquad + \delta(n-m+l+k)+\delta(n-m-l-k)+\delta(n-m+l-k)+\delta(n-m-l+k).
\end{eqnarray}
The zero mode has the canonical $3+1$ kinetic term with
field strength:   
\be
F_{\mu \nu}^{0~a} =
\partial_\mu A_{\nu}^{0~a} - \partial_{\nu}A_{\mu}^{0~a} + \tilde{g}
f^{abc} A_{\mu}^{0~b}A_{\nu}^{0~c},
\label{F4}
\ee
Hence, $\tilde{g}\equiv g/\sqrt{M_sR}$ is the dimensionless 
low-energy $3+1$ coupling
constant. If the truncation $N=M_sR$ on the number of the KK modes is 
introduced then $\tilde{g}\equiv g/\sqrt{N}$. A perturbative theory of the
zero mode requires $\tilde{g} < {\cal O}(1)$, i.e., $g< \sqrt{M_sR}$ or $M >
1/R$. 

\vskip 0.2cm

\noindent
{\em (iii) Comparison to Aliphatic Theory}
\vskip 0.1cm
Now,
consider again the
aliphatic theory with the gauge structure $SU(3)_{0}\times SU(3)_{1} \times
\dots \times SU(3)_{N}$, 
where the vector potentials are $ A_{\nu}^{j~a}$.  In addition, 
there are a set of $\Phi_{i}$ fields which 
straddle the $i$th and $i+1$th $SU(3)$ gauge groups. The
Lagrangian takes the form as in eqn.(1.1), and the mass spectrum
as in eqn.(1.3). 
The gauge fields $A_{\mu}^{j}$ can be expressed as linear combinations of the
mass eigenstates $\tilde{A}_{\mu}^{n}$ as: 
\be 
A_{\mu}^j = \sum_{n=0}^{N} a_{jn} \tilde{A}_{\mu}^n. 
\label{AA}
\ee
The $a_{nj}$ form a normalized
eigenvector ($\vec{a}_{n}$)
associated with the $n$th $n \neq 0$ eigenvalue 
and has the following components:
\be
a_{nj} =\sqrt{\frac{2}{N+1}} \cos{(\frac{2j+1}{2}\gamma_n})\ , \qquad
j = 0, 1,\dots, N, 
\label{anj}
\ee
The eigenvector for the zero-mode, $n=0$ ,
is always $\vec{a}_{0} =
\frac{1}{\sqrt{N+1}} (1,1,\dots,1)$. The orthogonality between the
eigenvectors is due to: 
\be
\sum_{j=0}^{N} \cos{(\frac{2j+1}{2}\gamma_n})  \cos{(\frac{2j+1}{2}\gamma_m})
= \delta(n-m) \frac{N+1}{2}\ , \qquad n,m \neq 0 \ll N
\ee
with $\gamma_{n} = \frac{n\pi}{N+1}$. 
We can now rewrite the Lagrangian eqn.(1.1) in the mass eigenstates of
the vector bosons ($\tilde{A}_{\mu}^{n}$) and derive the interactions between
them.

Let us now compare the KK reduction of the five-dimensional theory,
eqn.(\ref{L4}), and the aliphatic $(SU(3)^{N+1},\Phi^{N})$ 
theory at the level of interactions. 
In the aliphatic theory, as far as the mass spectrum 
is concerned, there are three free parameters, namely, the gauge coupling
constant $g_L$, the total number of $SU(3)$ groups $N+1$ and the VEV of the
Higgs field $v$. As we discussed earlier, one
can arrange the parameters of the $SU(3)^{N+1}$ 
theory to fix the ratio ${g_Lv}/{\sqrt{2}(N+1)} = \frac{1}{R}$, such that the
spacing of the linear mass spectrum at $n << N$ is completely
determined  and the mass spectrum of the two theories matches.

To compare the Lagrangian's couplings we substitute
eqn.(\ref{AA}) into the gauge part of the Lagrangian eq.(1.2): 
\begin{equation}
{\cal L}_{gauge} = -\frac{1}{4} \sum_{j=0}^{N}(
\sum_{n=0}^{N} a_{jn}\partial_{[\mu}A_{\nu]}^{n} + 
g_L f^{abc} \sum_{n=0}^{N}\sum_{m=0}^{N} a_{jn}a_{jm} A_{\mu}^{n}A_{\nu}^{m})^2
\end{equation}
Isolating the zero-mode, $\tilde{A}^{0}_\mu$, 
and, using orthonormality, we can write down the canonical kinetic terms: 
\begin{equation}
{\cal L}_{g, kin} =-\half(\partial_{[\mu}\tilde{A}_{\nu]}^{0} +
\frac{g_L}{\sqrt{N+1}}f^{abc}\tilde{A}_{\mu}^{0}\tilde{A}_{\nu}^{0})^2 +
\sum_{n=1}^{N}(\partial_{[\mu}\tilde{A}_{\nu]}^{n})^2. 
\end{equation}  
The trilinear gauge coupling takes the form:
\begin{equation}
{\cal L}_{g,3A} = -\frac{1}{4}\sum_{n,m,l\neq(0,0,0)} (\sum_{j=0}^{N}a_{jn}a_{jm}a_{jl}) g_L f^{abc}
\partial_{[\mu}\tilde{A}_{\nu ]}^{n} \tilde{A}^{m~\mu}\tilde{A}^{l~\nu}. 
\end{equation}
Using pairwise summations and orthogonality: 
\begin{equation}
\sum_{j=0}^{N}a_{jn}a_{jm}a_{jl} = \left\{ \begin{array} {cc}
	\sqrt{\frac{1}{N+1}}\left[\delta(n)\delta(m-l)+\delta(m)\delta(n-l)+\delta(l)\delta(n-m)\right] \ , & \\
	\sqrt{\frac{1}{2(N+1)}} \Delta_1(n,m,l)\ , & n,m,l \neq 0 
; 	
	\end{array} \right.
\label{Dnml}
\end{equation} 
where $\Delta_1$ is defined previously. 
Similarly, the quadrilinear
couplings take the form: 
\begin{equation}
{\cal L}_{g,4A} = -\frac{1}{4}\sum_{n,m,l,k\neq(0,0,0)}
(\sum_{j=0}^{N}a_{jn}a_{jm}a_{jl}a_{jk}) g_L f^{abc} g_L f^{ade}
\tilde{A}_{\mu}^{n} \tilde{A}_{\nu}^{m}\tilde{A}^{l~\mu}\tilde{A}^{k~\nu},  
\end{equation}
with the coefficients, 
\begin{equation}
\sum_{j=0}^{N}a_{jn}a_{jm}a_{jl}a_{jk} = \left\{ \begin{array} {cc}
	\frac{1}{N+1} , & {\rm two~ of~ (n,m,l,k)~ are~ zero, remainders~ are~equal}; \\
	\frac{1}{2(N+1)} \Delta_2(n,m,l,k)\ , & n,m,l,k \neq 0 
; 	
	\end{array} \right.
\end{equation} 
We see that 
$\Delta_2(n,m,l,k)$ is exactly 
the same function defined in the
discussion of truncated momentum space expansion. 
Thus, we see that, defining the gauge coupling constant
$\bar{g}=g_L/\sqrt{N+1}$ of the unbroken $SU(3)$ in the aliphatic theory to
satisfy $\bar{g}= \tilde{g} = {g}/\sqrt{M_s R}$, the couplings 
and Feynman rules in the two theories agree perfectly. 
This completes the demonstration of the equivalence.

In both theories, there
are three fundamental parameters, i.e., $M_s,~M=1/g_0^2,~R$ in the KK reduced theory 
 and $g_{L},~N,~v$ in the aliphatic theory. The mappings between them are $N+1 =
M_s R$, $g_L=\sqrt{M_s/M}$ and $v=\sqrt{M_sM}$, and they are valid
up to the scale $v$. Measurement of the zero mode
interactions give us $\bar{g}=\tilde{g}$. The mass of the first KK mode tells
us $\frac{g_L v}{N+1} = \frac{1}{R}$. Hence, two of the three parameters can
be determined, leaving $M_s = g_L v$ undetermined in the two theories. 
The mass of $M_2$ will test the linear spacing between the KK modes, rather
than give further constraints on the parameters.

Suppose we had a bulk $5+1$ theory. Then we would have a
different structure for the low energy effective theory,
and we would have a correspondingly different lattice
theory. No longer would the theory be an aliphatic model,
and would appear then as a more complex closed structure,
first an aromatic hydrocarbon, eventually a 
polymerized molecular solid state.
One can generalize our construction to theories in two extra-dimensions with
size $R_1 \times R_2$.  
The low energy effective theory would be different.

The simplest case is the limit of a single plaquette
in the two compact dimensions of $5+1$, 
the analogue of an Eguchi-Kawai model \cite{Eguchi}.
 The low energy theory would contain the gluon zero-mode, which is the
rotational zero-mode of such a configuration, and  
a doubly degenerate pair of colorons as the first KK modes,
and a third heavy singlet. One can expand the single plaquette construction to 
multi-plaquette construction, which requires $(N+1) \times (M+1)$ $SU(4)$ and
$2N \times M + N + M$ $\Phi_i$ fields, where $N= R_1/a_1$ and $M=R_2/a_2$ and
$a_1, ~a_2$ are spacings between the 3-branes.  

It is interesting that ultimately the lattice structure
must also reflect the homotopy of the extra dimensions.
If there is a ``hole'' in the space of the extra
dimensions, there must be corresponding nontrivial
paths through the Higgs field links that match the 
non-contractable loops in that space.

\begin{figure}
\vskip -1in
\centerline{
\hbox{
\epsfxsize=2.5truein
\epsfbox[70 32 545 740]{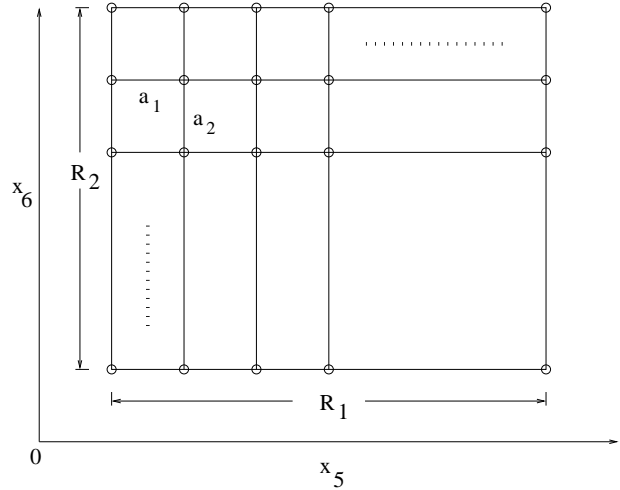}
}
}
\caption{The geometric interpretation for the plaquette model with two extra
dimensions. The Eguchi-Kawai model
corresponds to a single plaquette. 
At each circle, there is a 3-brane with one $SU(3)$ symmetry. }  
\end{figure}

\noindent
\section{Incorporation of Fermions}

The models we presented for the gauge bosons in the bulk can easily
accommodate fermions and bosons in the bulk. 

The Lagrangian for a fermion in the five dimensional bulk which is charged
under the bulk $SU(3)$ symmetry is given by 
\be
{\cal L}_{5}(x^{\mu}, x^5) = \bar{\Psi}(i\gamma^{\mu}D_{\mu} - \gamma_5 D_{5}) 
\Psi - \frac{1}{4} Tr(F^{MN}F_{MN}), 
\ee
where the covariant derivative is defined previously. 
The five dimensional fermion is non-chiral, hence its zero mode upon 
the compactification of the 5th dimension can be non-chiral, unless the
Lorentz group in five dimensions $SO(4,1)$ is explicitly broken by imposing
different boundary conditions for the left-handed component, $\Psi_L$, and
the right-handed component, $\Psi_R$. The boundary conditions also prevents
 $\Psi$ from having a bare mass term in the bulk. Consider, for example, the
following boundary condition, 
\be
\label{bc}
\frac{\partial}{\partial x_5}\Psi_{L} |_{x_5=0, R} = 0; \qquad
\Psi_{R}|_{x_5=0,R} = 0.  
\ee
The Neumann boundary condition for $\Psi_L$ ensures that there is a massless
left-handed four dimensional fermion on the brane, while the Dirichlet
conditions makes all the right-handed modes massive. Upon compactification,
$\Psi_{L}$ can be decomposed into a cosine series and $\Psi_{R}$ can be
decomposed into a sine series. The masses of the fermion KK modes are given
by $M_{L/R, n} = n \pi /R$.  

In the aliphatic model, consider $N+1$ fermions $\Psi_{n}$ ($n=0 \cdots N$), each of which is charged under the corresponding
$SU(3)_n$ symmetry. The Higgs fields $\Phi_{n}$ which is $(\bar{3},3)$ under
the two neighboring $SU(3)$ symmetries provides the nearest neighbor couplings 
between the fermion fields. The effective Lagrangian takes the form 
\be
{\cal L}_{fermion}(x^{\mu}) =  \sum_{n=0}^{N} \bar{\Psi}_{n,L/R}\slash{D}\Psi_{n,L/R}
+ M_f \left[\bar{\Psi_{n,L}}(\frac{\Phi_{n+1}^{\dagger}}{v}\Psi_{n+1, R}
-\Psi_{n,R}) - \bar{\Psi_{n,R}}(\Psi_{n, L}- \frac{\Phi_{n}}{v}\Psi_{n-1,L})\right],  
\ee
where $\slash{D}$ is defined as the four dimensional covariant derivative. 

In the aliphatic model, the
boundary conditions in eqn. (\ref{bc}) can be translated into $\Psi_{0,R} = \Psi_{N, R}=0$ and
$\Psi_{L,N}-\Psi_{L,N-1}=0$. As a result, in the vacuum
where $\Phi_{n}$ has non-zero VEV $v$, the mixed  mass terms for the
left-handed and right-handed fermions are  
\be
\begin{array}{ccc}
{\cal L}_{mass} & = & M_f\left\{ \bar{\Psi_{0,L}}\Psi_{1,R} + \sum_{n=1}^{N-1}
\left[ \bar{\Psi_{n,L}}(\Psi_{n+1,R}-\Psi_{n,R})-
\bar{\Psi_{n,R}}(\Psi_{n,L}-\Psi_{n-1,L}) \right] \right\} \\
& = & (\bar{\Psi_{0,L}}, \cdots, \bar{\Psi_{N-1,L}}) M (\Psi_{1,R}, \cdots,
\Psi_{N-1,R})^{T};  
\end{array}
\ee
where the $N \times (N-1)$ mass matrix $M$ takes the form 
\be
M = M_f \left(
\begin{array}{cccc}
1&0&\cdots&0 \\
-1&1& \cdots&0 \\
& &\cdots& \\
0&\cdots&-1&1 \\
0&\cdots&0&-1
\end{array} \right).
\ee

To calculate the mass eigenvalues and eigenstates for the right-handed
components, one can diagonalize the $(N-1)\times (N-1)$ matrix $M^{\dagger}M$, 
\be
M^{\dagger}M = |M_f|^2 \left(
\begin{array}{ccccc}
2&-1&0&\cdots&0 \\
-1&2&-1& \cdots&0 \\
0&-1 &2 &\cdots&0 \\
& & & \cdots & \\
0&0&\cdots&-1&2
\end{array} \right).
\ee
Therefore, the eigenvalues of the right-handed fermions are
\be
M_{R,n} = 2 M_f \sin{(\frac{n\pi}{2N})}, \qquad n=1,2, \cdots, N-1.
\ee  
In terms of the mass eigenstates $\tilde{\Psi}_{n,R}$, 
\be 
\Psi_{n,R} = \sqrt{\frac{N}{2}}\sum_{k=1}^{N-1} \sin{(n\frac{k\pi}{N})} \tilde{\Psi}_{k,R}. 
\label{FR}
\ee

The mass eigenvalues of the left-handed fermions can be calculated from the $N
\times N$ matrix $M M^{\dagger}$, which takes the following form, 
\be
MM^{\dagger} = |M_f|^2 \left(
\begin{array}{ccccc}
1&-1&0&\cdots&0 \\
-1&2&-1& \cdots&0 \\
0&-1 &2 &\cdots&0 \\
& & & \cdots & \\
0&0&\cdots&-1&1
\end{array} \right).
\ee
Hence, the eigenvalues of the
left-handed fermions are similar to those of the gauge bosons, 
\be
M_{n,L}= 2 M_f \sin{(\frac{n\pi}{2N})},  \quad n=0 \cdots N-1. 
\ee
Hence, the left-handed fermions have a massless zero mode. The massive modes
have the same mass as those of the right-handed fermions, thus form massive
vector pairs. 

The eigenvectors of the left-handed fermions also have the same structure as that of
the gauge bosons, namely, in terms of the mass eigenstates $\tilde{\Psi}_{k,L}$,
\be 
\Psi_{n,L} = \sqrt{\frac{N}{2}}\sum_{k=0}^{N-1} \cos{(\frac{2n+1}{2}\frac{k\pi}{N})} \tilde{\Psi}_{k,L}. 
\label{FL}
\ee
Note that left-handed fermions have a $\cos$ expansion, while the right-handed
fermions assume a $\sin$ expansion. 

In the limit that $n \ll N$, a linear massive spectrum is recovered for both
right-handed and left-handed fermions, in which
$M_n = M_f \frac{n \pi}{N}$. Since the masses of the KK
modes for a $D=5$ fermion are $M_{L/R,n} = \frac{n \pi}{R}$,  one 
reproduces the linear spectrum for the KK theory by choosing $M_f = \frac{N}{R}$.

The coupling
between the fermions and the gauge field takes the following form in their
mass eigenstate basis, 
\be
\begin{array}{ccc}
{\cal L}_{ffA} & = &  \sum_{n,m,l \neq (0,0,0)} g_{L}
\tilde{\bar{\Psi}}_{n,L} \gamma^{\mu} \tilde{A}_{\mu m} \tilde{\Psi}_{l,L}
\Delta_{n,m,l} + g_{L} \tilde{\bar{\Psi}}_{0,L} \gamma^{\mu} \tilde{A}_{\mu 0}
\tilde{\Psi}_{0,L} \\
& & + \sum_{n,m,l \neq 0, N} g_{L} \tilde{\bar{\Psi}}_{n,R} \gamma^{\mu}
\tilde{A}_{\mu m} \tilde{\Psi}_{l,R}\Delta_1(n,m,l)  , 
\end{array}
\ee
in which $\Delta_1$ is defined as the sum in eqn.(\ref{Dnml}).

One can also write down the effective Lagrangian for a massless complex boson
in the bulk in our frame work. Consider $N+1$ 4D complex scalar with the
following Lagrangian,    
\be
{\cal L}_{boson} = \sum_{i=0}^{N} |D_{\mu}\phi_{i}|^2 - M_b^2 \sum_{i=1}^{N}
|\phi_{i-1}-\frac{1}{v}\Phi_{i}\phi_{i}|^2. 
\ee 
In the vacuum in which $\langle \Phi_{i} \rangle = v$, the scalars have the
mass terms $-M_b^2 \sum_{i=1}^{N} |\phi_{i-1}-\phi_{i}|^2$. It can
diagonalized by 
\be
\phi_{j} = \frac{1}{N+1} \sum_{n=1}^{N} e^{i 2\pi nj/(N+1)} \tilde{\phi}_{n}, 
\ee
with the mass spectrum 
\be
M_{n,b}= 2M_b \sin{\gamma_n}, \quad n=0, 1, \cdots N. 
\ee
Each level with $n\neq 1$ is degenerate with the level $N-n$, while the zero
mode is a singlet. This doubling of energy levels corresponds to the mode
expansion in $x^5$ in terms of $1$, $\sin(n \pi x^5/R)$ and $\cos(n \pi
x^5/R)$, where the sine and cosine terms are degenerate modes.

\noindent
\section{Renormalization of gauge coupling constant }

Unlike the compactified continuum theory, the spontaneously broken gauge
theory $(SU(3)^{N+1}$ $,~\Phi^{N})$ is a renormalizable field theory. Thus, we
can discuss the scale dependence of the 
coupling strength $\bar{g}(\mu)$ of the unbroken $SU(3)$ via the radiative
corrections. The standard method of constructing effective field theories at
each stage of the decoupling of the massive modes is at best confusing. One
problem is that when decoupling the $n_{th}$ KK mode with mass $M_{n}$, the
decoupling methods tells us to construct an effective theory with one zero mode 
and $n-1$ KK modes which should be taken to be massless at the decoupling
scale $M_n$, this is, the effective field theory will have a gauge symmetry
$SU(3)^{n}$. But the original theory tells us that all $SU(3)^{N+1}$ is broken 
to $SU(3)$ at the scale $v$, and it is different from breaking the $SU(3)$
symmetries one by one at each $M_{n}$. Another problem is that, at two or
higher loop level, one necessarily encounters loops with both light and heavy
KK modes, such that it is confusing to even define a proper decoupling scale.  

However, one can define the effective coupling constant $\bar{g}(\mu ^2)$ in
the momentum subtraction scheme \cite{PQW}, e.g., as the triple gluon (zero
mode) vertex. All the external legs have the momentum $q^2 = - \mu ^2$. The
effective coupling $\bar{g}(\mu ^2)$ is governed by the equation
\be
\label{beta}
\frac{\partial{\bar{g}(\mu ^2)}}{\partial{\ln \mu ^2}} =  \beta(\bar{g}(\mu
^2)),
\ee
and its evolution can be calculated in any order of perturbation in the full
spontaneously broken $(SU(3)^{N+1},~\Phi^{N})$ theory, including all KK modes.
Strictly speaking, one gets a set of coupled differential equations, since the 
$\beta$ function in eqn. (\ref{beta}) depends on the triple vector boson
couplings $\bar{g}_{0nn}(\mu ^2)$, each running according to its own evolution 
equation. The problem radically simplifies at 1-loop level and in the
approximation \cite{PQW} where one assumes that the KK modes that appears in
the loops satisfy $-\mu^2 \leq (M_{i}^2 + M_{j}^2)$. Moreover, in 1-loop
calculation one can use the relationship between tree level couplings, namely, 
$\bar{g} = \bar{g}_{0nn}$ for any $n$. 

Thus, at 1-loop level,
the running of the gauge coupling constant $\bar{g}$ between the scales
($M_n$, $M_{n-1}$) involves the $n$ modes which are lighter
then $M_n$, as a result the running can be described by, 
\begin{equation}
\frac{d\bar{g}}{d\log \mu}= - [n \frac{\beta}{4\pi^2}]~ \bar{g}^3, \qquad
M_{n-1} \leq \mu \leq M_{n}; 
\end{equation}   
in which $\beta$ is the 1-loop RGE coefficient of a pure $SU(3)$
theory. Hence, given the measured coupling constant $\alpha(M_Z)$ at low
energy, the gauge coupling constant at energy scale $\mu$ is given by  
\be
\alpha^{-1}(\mu) = \alpha^{-1}(M_Z) -
\frac{\beta}{4\pi}\left[\ln(\frac{M_1}{M_Z}) + \sum_{n=2}^{n_{max}} n
\ln(\frac{M_n}{M_{n-1}}) +
(n_{max}+1)\ln(\frac{\mu}{M_{n_{max}}})\right], 
\ee
where $M_{n_{max}} \leq \mu < M_{n_{max}+1}$. One can sum up the series to
arrive at, 
\be
\label{alpha}
\alpha^{-1}(\mu) = \alpha^{-1}(M_Z) - \frac{\beta}{4\pi}\ln(\frac{\mu}{M_Z}) 
- \frac{\beta}{4\pi}  n_{max} \ln(\frac{\mu}{M_1}) + \frac{\beta}{4\pi} F(M_n), 
\ee
in which the factor $F\equiv  \ln \left( \frac{\Pi_{n=1}^{n_{max}}
M_n}{M_1^{n_max}} \right)$ depends on what kind of the KK spectrum we work
with. The linear spaced KK spectrum from the dimensionally reduced continuum
theory gives:
\be
F_{lin} = \ln(n_{max}!);
\ee
while the spectrum from the aliphatic model as in eqn. (\ref{Mn}) gives:
\be
F_{ali} = \ln \left( \frac{\Pi_{n=1}^{n_{max}}
\sin(\frac{n\pi}{2(N+1)})}{\sin(\frac{\pi}{2(N+1)})^{n_{max}}} \right). 
\ee
Eqn. (\ref{alpha}) with $F_{lin}$ is derived in \cite{dienes} and
\cite{dobrescu}, it shows a power law behavior of the gauge coupling constant. 
The differences between $F_{lin}$ and $F_{ali}$ provides an interesting
measure on how much the aliphatic mode deviates from the continuum theory at a 
quantum level. In Fig. (3), we plot $F_{lin}$ and $F_{ali}$ as a function of
$n_{max}$, keeping $N$ fixed. Fig. (4) shows $F_{lin}$ and $F_{ali}$ as a
function of $N$, while $n_{max}$ is fixed. 
\begin{figure}
\centerline{
\hbox{
\epsfxsize=2.5truein
\epsfbox[70 32 545 740]{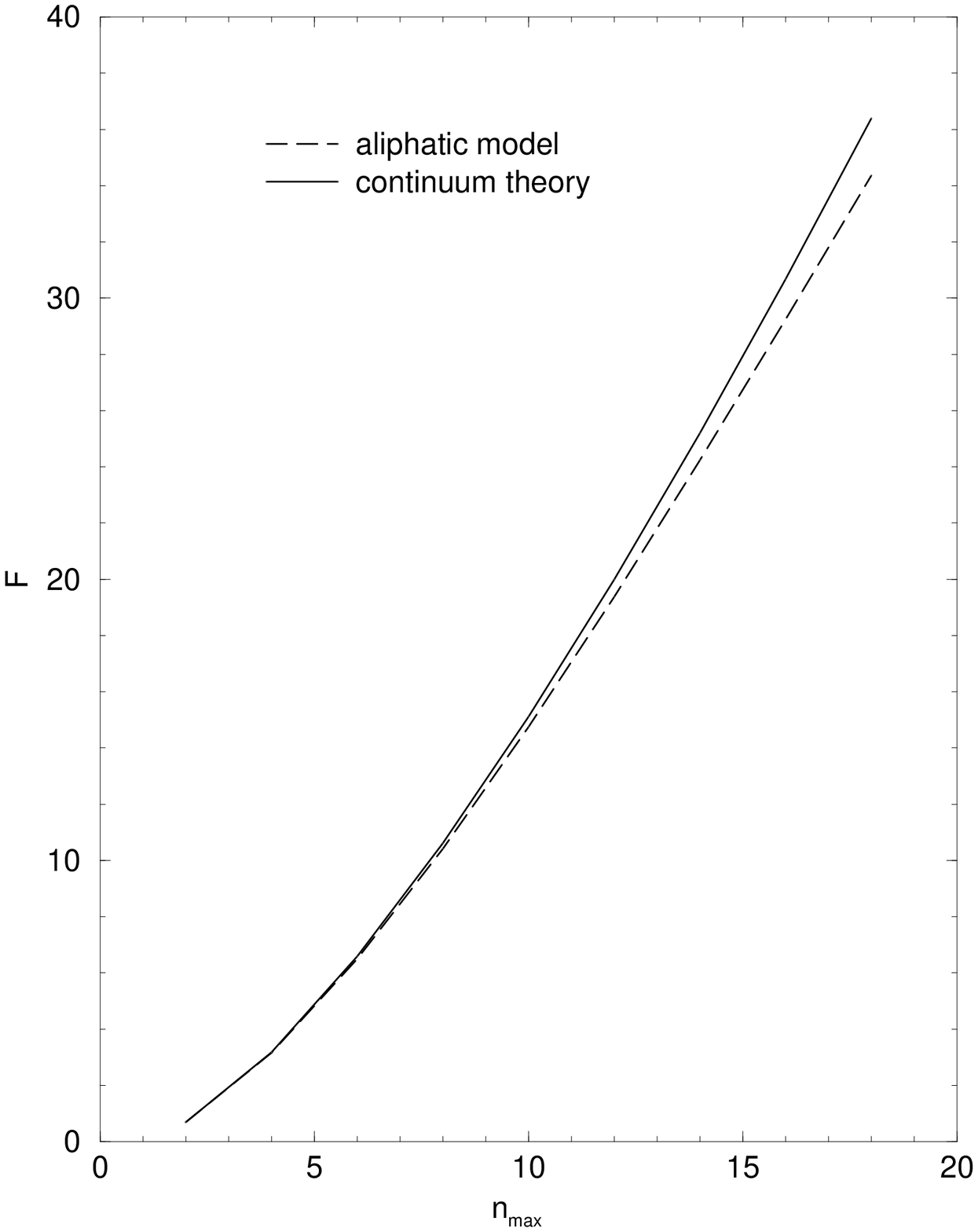}
}
}
\caption{$F_{lin}$ (solid line) and $F_{ali}$ (dashed line)
as functions of $n_{max}$. $N=20$ is chosen for the plot.}

\centerline{
\hbox{
\epsfxsize=2.5truein
\epsfbox[70 32 545 740]{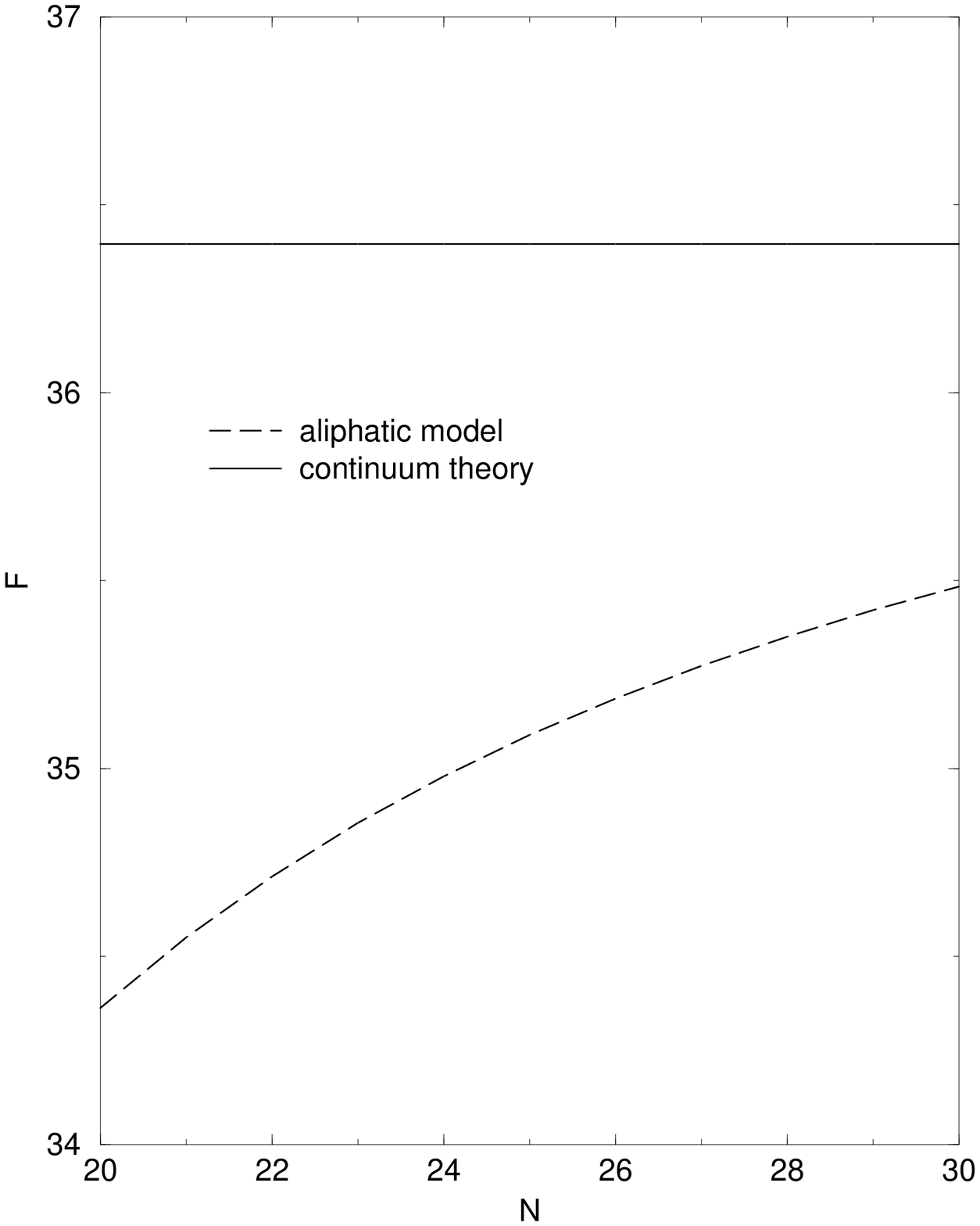}
}
}
\caption{$F_{lin}$ (solid line) and $F_{ali}$ (dashed line)
as functions of $N$. $n_{max}=18$ is chosen for the plot.}  
\end{figure}
  
It can be seen from the figures that when $n_{max}$ is small compared to $N$,
the two theories agree very well in their $\beta$ functions, since in this
region, the aliphatic model gives an excellent approximation of the linear
spectrum from the compactified continuum theory, as we expected. 
However, even at $n_{max}$ close to $N$, the
deviation of $F_{ali}$ from $F_{lin}$ is less than $10\%$. 

We don't expect
things will be drastically different at two or higher loop levels. This
observation suggests to us that the aliphatic model
provides a good approximation to the continuum theory even at an energy scale
close to $v$, the ``error'' in approximating the continuum theory lies in the
finite size of the lattice, i.e., the separation between the two nearest
branes. One can always reduce the ``error'' by adding more branes, thus
increasing $N$ and reducing the inter-brane separation. It also suggests to us 
that, if one wants to modify the aliphatic theory such that it will produce
exactly the linear spectrum up to $M_{N}$, one only needs to add higher order
operators, perhaps the type of operators which mimic the couplings between 
the next-to-nearest branes.

\section{Discussion and Conclusion}

In eq.(2.3)
we assume that one can lift the Higgs
mass to a high energy
scale above the cut-off scale $M_s $.
The Higgs degrees of freedom  then decouple from the theory,
and only the Nambu-Goldstone modes remain, which are eaten
by the KK modes to give mass.  This is a large coupling
limit of the Higgs theory in which the 
VEV is held fixed, i.e., $v \sim M/\sqrt{\lambda}$ where
$M\rightarrow \infty$ and $\lambda\rightarrow \infty$
together.  However, such a theory violates perturbative
unitarity. On the other hand, the effective
low energy theory is a gauged chiral Lagrangian
with $f_\pi\sim v$.  This theory is a pertubatively
sensible one (and is renormalizeable
as expansion in $1/v^p$) in the low energy limit,
however, the perturbative unitarity breakdown occurs
when $\sqrt{s} \gta v$. 
Essentially, longitudinal KK mode
scattering must violate perturbative unitarity when
${s} \gta 4\pi v^2$. This is 
the Lee-Quigg-Thacker bound which applies to, e.g., electroweak 
symmetry breaking for $WW$ scattering \cite{Lee}.  

We see, from eq.(2.7), that this failure
of unitarity 
corresponds to energy scales approaching 
${s} \gta 4\pi N^2/g^2_L R^2
\sim 4\pi N^2M/R^2M_s $.  As we have seen,
our theory corresponds to a $4+1$
theory with a dimensional coupling
given by $g_0$. We would generally expect
this theory to violate perturbative
unitarity for ${s} \gta 4\pi M_s/g_0^2$, 
hence, by comparison that indeed
${s} \gta 4\pi N^2M/R^2M_s \sim 4\pi M_s/g_0^2$.
Hence the perturbative unitarity violation inherent in the
large coupling constant of the parent $D=5$ theory
is matched by the unitarity breakdown in the effective
$3+1$ theory. 

The separation of scales, $N\sim M_s/M_c>>1$ is a requirement of
very low mass, or infrared states, in an
essentially strong-dynamical theory at the scale $M_s$,
In all cases in nature where this phenomenon
occurs and is understood, there is an attendant
custodial symmetry.   
The theory we have presented in $3+1$ dimensions
imitates arbitrarily well a $4+1$ theory, 
and this dynamical issue does not seem
to arise.  The infrared physics scale,
the  ``effective
compactification scale,'' is $M_c \sim M_s/N$
and apparently occurs accidentally because $N$, the number
of independent gauge groups in the contruction, 
is very large. 

One might have thought that
the separation of the compactification scale and
the fundamental scale in extra-dimensional models
would involve, at least accidentally, 
approximate classical scale invariance
(this is the custodial symmetry in QCD of, e.g., the
ratio $\Lambda_{QCD}/M_{Planck}$ in the sense that
``classical scale invariance''
corresponds to setting the $\beta$-function of QCD to zero).
The QCD coupling in our theory
turns out to
be suppressed as $\alpha_{QCD} \sim M_s \alpha_0/N $,
where $\alpha_0 =g_0^2/4\pi $ is the dimensional
$4+1$ gauge coupling.
To take
$N$ arbitrarily large thus
implies that the theory must have a
slowly running dimensionless couplng constant
(remniscent of ``walking technicolor'') in $D=4$ on scales
well below $M_s$, so it does appear that quantum scale breaking
effects are under control, and it seem that classical
scale invariance is acting as the custodial
symmetry afterall. 
However, the trace of the stress-tensor in $D=5$ is
nonzero classically, and the theory has 
explicit scale breaking, owing
to the $D=5$ dimensional coupling constant.   The nonzero trace,
$T_\mu^\mu \propto G^a_{\mu\nu}G^a{}^{\mu\nu}$ in $D=5$
must match onto the KK masses as in $D=4$,
since the KK masses are seen as explicit
sources of scale breaking on all scale from $M_c$
to $M_s$. It is therefore quite
puzzling as to 
what, if anything, we may we invoke as 
the custodial symmetry of the scale hierarchy
in extra dimensions when $N$ is large.  
Is this a counter example to the requirememnt
of having an explicit custodial symmetry, an
artifact of large $N$?

In conclusion, We have constructed a manifestly gauge invariant description
of $n$ KK modes for an $SU(m)$ gauge theory in the bulk.
We showed in this paper the four-dimensional KK
theory deducted from a compactified five-dimensional $SU(3)$ theory can be
considered as a 
$(SU(3)^{N+1},~\Phi^N)$ theory, in which the $SU(3)^{N+1}$ gauge symmetry is
spontaneously broken to $SU(3)$. This theory owes its structure to a
transverse lattice theory with one extra dimension. The three 
dimensional parameters of the original KK theory, the string (cut-off) scale
$M_s $, the compactification radius $R$ and the five-dimensional gauge 
coupling $g_0 \equiv \sqrt{M^{-1}}$, determine the structure of the
$(SU(3)^{N+1},~\Phi^{N})$ theory: $N=M_sR$, the coupling constant of the
unbroken $SU(3)$ $\bar{g}= 1/\sqrt{MR}$, and the scale $v=\sqrt{M_sM}$ of the
spontaneous symmetry breaking $SU(3)^{N+1}\rightarrow SU(3)$. 

The approach maintains manifest gauge invariance.
Is it possible to construct analogous effective
Lagrangians which maintain SUSY and general
covariance for yielding KK modes of gravity?
And how are the topological aspects of 
extra dimensional gauge theories \cite{ramond}
expressed in an effective Lagrangian
such as this?

(Note added:) Upon completion of this work
the preprint of Arkani-Hamed, Cohen and Georgi, \cite{georgi}, 
appeared which uses a technicolor-like
condensate in place of our explicit Higgs
fields, $\Phi_n$, but obtains essentially the
identical construction as a chiral Lagrangian.  
Georgi's moose
notation, used in \cite{georgi}, may be 
a useful way to extend to higher dimensions
such as $5+1$ with $2$ compact dimensions, whence
the theory may be graphically
represented as a ``moose lattice,'' and
the anomaly free incorporation of fermions 
is automatic.\\

\noindent
{\bf \Large \bf Acknowledgements}

We wish to thank W. Bardeen, Hsin Chia Cheng
and Martin Schmaltz for useful discussions.
One of us (SP) wishes to thank P. Chankowski and M. Misiak for useful
discussions and acknowledge the kind hospitality of the
Fermilab theory group where this work originated.
Research by CH and JW was supported by the U.S.~Department of Energy
Grant DE-AC02-76CHO3000.
Research by SP was supported by the Polish State Committee
for Scientific Research, grant KBN 2 P03B 060 18
(2000-01).

\end{document}